# Stereotactic Arrhythmia Radioablation for Refractory Ventricular Tachycardia: A Narrative Review and Exploratory Pooled Analysis of Clinical Outcomes and Toxicity

Keyur D. Shah, PhD[1], Chih-Wei Chang, PhD[1], Sibo Tian, MD[1], Pretesh Patel, MD[1], Richard Qiu, PhD[1], Justin Roper, PhD[1], Jun Zhou, PhD[1], Zhen Tian, PhD[2] and Xiaofeng Yang, PhD[1*]

[1]Department of Radiation Oncology and Winship Cancer Institute, Emory University, Atlanta, GA

[2]Department of Radiation & Cellular Oncology, University of Chicago, Chicago, IL

*Corresponding to: xiaofeng.yang@emory.edu

**Running title:** STAR for Ventricular Tachycardia: Narrative Review & Pooled Analysis

**Manuscript Type:** Review Article

**Funding:** This research is supported in part by the National Institutes of Health under Award Number R01CA272991, R01EB032680, R37CA272755 and U54CA274513.

**Declaration of interest:** None

## Abstract

**Purpose:** Stereotactic arrhythmia radioablation (STAR) is a non-invasive salvage therapy for refractory ventricular tachycardia (VT), especially in patients ineligible for catheter ablation. This narrative review and pooled analysis evaluates the safety, efficacy, and technical characteristics of STAR, integrating preclinical studies, case reports, case series, and clinical trials.

**Methods and Materials:** A comprehensive review identified 86 studies published between 2015 and 2025, including 12 preclinical studies, 49 case reports, 18 case series, and 7 clinical trials. Study-level data were extracted for pooled analysis of 6- and 12-month mortality, VT burden reduction, and grade 3+ acute toxicities. Subgroup analyses were performed by delivery modality, age, left ventricular ejection fraction (LVEF), and cardiomyopathy type.

**Results:** Pooled mortality was 16% (95% CI: 11–20%) at 6 months and 33% (95% CI: 27–38%) at 12 months. VT burden reduction at 6 months averaged 75% (95% CI: 73–77%) but showed substantial heterogeneity (I² = 98.8%). Grade 3+ acute toxicities occurred in 7% (95% CI: 4–10%), with heart failure being most common. Subgroup analyses suggested better outcomes in younger patients, those with NICM, and those with higher LVEF.

**Conclusions:** STAR is a promising salvage therapy with favorable acute safety and efficacy. Outcome heterogeneity and inconsistent reporting highlight the need for standardized definitions, dosimetric protocols, and longer-term follow-up. Prospective trials and real-world registries are critical for refining STAR's role in VT management.

# 1. Introduction

Ventricular tachycardia (VT) is a life-threatening arrhythmia characterized by rapid electrical activity originating in the ventricles, which can compromise cardiac output and lead to hemodynamic instability [1,2]. VT circuits are commonly associated with myocardial scar tissue, typically resulting from conditions such as myocardial infarction or cardiomyopathies [3,4]. The reentrant mechanism is a prevalent cause of VT, where electrical impulses circulate through areas of slow conduction, including the central isthmus, entrance, and exit sites, which sustain arrhythmic activity [5,6]. Between 2007 and 2020, ventricular tachycardia (VT) was linked to over 7,000 deaths in the United States among patients with underlying cardiovascular disease, underscoring the significant mortality burden associated with this arrhythmia [7].

The management of VT includes pharmacological therapy, device-based interventions, and catheter ablation [8]. Antiarrhythmic drugs such as amiodarone and beta-blockers are first-line treatments, but their efficacy is often limited, and they may carry significant side effects [9]. Implantable cardioverter-defibrillators (ICDs) are a cornerstone in preventing sudden cardiac death by terminating VT episodes through either anti-tachycardia pacing or defibrillation shocks [10]; however, they do not prevent arrhythmia recurrence and can negatively impact the quality of life due to frequent shocks and complications [11].

Catheter ablation has emerged as a primary interventional strategy, aiming to eliminate arrhythmogenic substrates by using radiofrequency or cryothermal energy to create lesions that disrupt reentrant circuits [12]. Despite its effectiveness, catheter ablation is invasive and may be challenging in patients with extensive myocardial scar burden or hemodynamic instability. Procedural success is highly dependent on precise identification and targeting of VT circuits, often guided by electroanatomical mapping (EAM), which integrates functional and anatomical data for improved precision. While catheter ablation is a primary interventional strategy for VT, it is not curative in all cases, particularly in patients with extensive scar burden or inaccessible arrhythmogenic tissue. Additionally, the procedure is associated with elevated risks of complications and mortality, especially in those with advanced heart failure or significant comorbidities [13,14].

Stereotactic arrhythmia radioablation (STAR) has emerged as a promising non-invasive salvage option for treating VT in patients who are not candidates for or have failed catheter ablation. STAR offers a non-invasive alternative for VT treatment, leveraging high-dose radiation to modify arrhythmogenic substrates. Unlike catheter-based interventions, STAR eliminates the risks of invasive procedures while providing precision targeting through advanced imaging modalities [15]. The concept of using radiation therapy for VT management dates to the early 2000s in Japan. Miyashita et al [16] pioneered this approach by combining chemotherapy and RT to treat a case of right ventricular outflow tract (RVOT) VT in a 70-year-old female, delivering 40 Gy in a single fraction. This early effort was followed by Tanaka et al [17], who applied a similar chemo-RT regimen with 51 Gy in a single fraction to manage RVOT VT in a 65-year-old male. While these exploratory studies highlighted the potential of RT to address arrhythmic substrates, the lack of advanced imaging and delivery techniques limited the precision and safety of these early treatments.

The modern era of STAR began in 2015 when Loo et al [18] at Stanford demonstrated the first in-human STAR procedure as it is known today. Using the CyberKnife system, they delivered 25 Gy in a single fraction to treat VT and achieved a remarkable 90.7% reduction in arrhythmic events. This groundbreaking work established the feasibility of non-invasive VT ablation and paved the way for further exploration. In 2017, Cuculich et al [19] reported the first case series of STAR, treating five patients with 25 Gy in a single fraction and demonstrating an impressive 99.99% reduction in VT events. This study marked a pivotal moment in the field, showcasing STAR's clinical efficacy and safety.

Utilizing advanced imaging modalities and precision delivery platforms, STAR delivers highly conformal radiation doses to arrhythmogenic substrates. Its key advantages include non-invasive delivery, outpatient feasibility, and the ability to target arrhythmogenic foci inaccessible by catheter-based approaches. The growing body of evidence supports STAR as an effective option for reducing VT burden, with favorable acute and mid-term outcomes. Most STAR treatments prescribe a dose of 25 Gy in a single fraction, guided by preclinical studies and clinical experience, with delivery platforms such as linear accelerators (LINACs), CyberKnife, and MRI-guided systems offering unique capabilities in motion management and precision targeting.

Given the increasing adoption of STAR for VT management, there is a need for a comprehensive synthesis of the available evidence to assess its safety and efficacy. Several recent systematic reviews have summarized STAR outcomes, focusing on overall mortality, VT burden reduction, and early toxicity profiles [20–23]. However, these studies have not evaluated outcome heterogeneity across clinical subgroups, such as delivery modality, ejection fraction, age, or underlying cardiomyopathy type. This narrative review and pooled analysis address this gap by presenting the first exploratory subgroup synthesis of STAR outcomes, using available study-level data to investigate how key clinical and technical variables may influence treatment response. This narrative review and pooled analysis aims to:

- Summarize key findings from preclinical studies and case reports to provide translational and mechanistic context for STAR.
- Quantify mortality rates at 6- and 12-months following STAR for VT in clinical cohorts.
- Assess the efficacy of STAR in reducing VT burden, including subgroup analyses based on delivery modality, patient characteristics, and cardiomyopathy type.
- Evaluate the incidence of acute grade 3+ toxicities associated with STAR in clinical studies.

This narrative review and pooled analysis includes data from preclinical studies, case reports, case series, and clinical trials, offering a comprehensive perspective on the evolving role of STAR in VT management. The findings aim to guide clinical practice and inform future research efforts to optimize patient selection, treatment planning, and post-treatment surveillance.

## 2. Methods

**2.1 Search Strategy and Study Selection**

A comprehensive literature search was conducted to identify relevant studies evaluating the use of radiation therapy for VT. The search was performed in PubMed, using the search term: "ventricular tachycardia AND radiation therapy." The search covered studies published up to May 13, 2025, with filters applied to include specific study designs such as preclinical investigations, case reports, case series, and clinical trials. A total of 350 studies were initially identified through the database search. After the removal of duplicates, the remaining studies underwent screening based on title and abstract, followed by full-text review.

Inclusion criteria were as follows:

- Studies reporting on preclinical investigations, case reports, case series, and clinical trials.
- Studies published in English.
- Studies reporting on distinct patient populations; in cases where multiple studies reported on the same cohort, the most recent publication was included.

Studies were excluded if they:

- Insufficient reporting on VT treatment using radiation therapy.
- Studies focused on unrelated topics, lacking clinical or preclinical relevance to radiation therapy for VT.
- Review articles, editorials, or commentaries that did not report original data.

Given the narrative and exploratory nature of this review, the search strategy was designed for breadth and relevance rather than exhaustive systematic retrieval. Figure 1 illustrates the study selection process, detailing the number of records screened, included, and excluded at each stage.

**2.2 Data Extraction and Analysis**

Data were systematically extracted from the included studies, focusing on key parameters related to patient demographics, treatment characteristics, and clinical outcomes. The following variables were collected:

- Patient Characteristics: Sample size, median age, gender distribution, underlying cardiomyopathy (ischemic vs. non-ischemic), median left ventricular ejection fraction (LVEF) and prior catheter ablation history.
- Treatment Parameters: Radiation delivery modality (LINAC, CyberKnife, or MRI-guided systems), prescribed dose, and planning target volume (PTV) margins.
- Clinical Outcomes: 6- and 12-month mortality rates, VT burden reduction, adverse events (grade 3+ toxicity) within 90 days.

Adverse events were classified according to the Common Terminology Criteria for Adverse Events (CTCAE), where reported. For studies that did not specify the grading system, adverse events labeled as grade 3 or higher by the authors were included. When data were missing or ambiguously reported, the studies were excluded from the pooled analysis for that specific variable to maintain the robustness of results.

### 2.3 Statistical Analysis

The pooled analysis was performed using a random-effects model to account for potential variability across the included clinical studies (case series and clinical trials). Key outcomes analyzed included mortality rates, efficacy measures, and safety profiles. Mortality rates were evaluated at 6 and 12 months. Subgroup analyses were conducted to examine differences based on cardiomyopathy type (ischemic vs. non-ischemic), patient age (≤ median vs. > median), and LVEF (≤ median vs. > median). Age and LVEF subgrouping was performed using the median values reported at the study level. For each study, if the cohort-level median (or mean, where median was unavailable) was above or below the pooled median across all studies, it was classified accordingly. For cardiomyopathy, patients were stratified based on studies that clearly separated ischemic (ICM) and non-ischemic (NICM) etiologies; studies that reported mixed or ambiguous labeling were excluded from this subgroup analysis. Safety outcomes were reported as rates of grade 3 or higher adverse events occurring within 90 days of treatment. Treatment efficacy was assessed through pooled analysis of VT burden reduction, quantified as the percentage reduction in VT events at 6 months. Only studies that reported or enabled calculation of VT burden reduction over a 6-month interval were included for this endpoint. VT burden was defined at the study level and included ICD-treated VT episodes, sustained VT, or total arrhythmic events. Most estimates were derived from study-level summaries (e.g., group means or medians); patient-level data were not available. Blanking periods were variably applied and not consistently reported across studies, contributing to outcome heterogeneity.

Heterogeneity across studies was assessed using the I-squared ($I^2$) statistic and Cochran's Q test. $I^2$ values of 50% or higher were interpreted as indicative of moderate-to-high heterogeneity, and potential sources of heterogeneity were further explored through subgroup analyses. These subgroup analyses included comparisons between LINAC- and CyberKnife-based treatments and evaluations of the impact of patient demographics, such as age and LVEF, on treatment outcomes. All statistical analyses were performed using Python, utilizing the Statsmodels and SciPy libraries. Results were presented as pooled estimates with corresponding 95% confidence intervals (CIs).

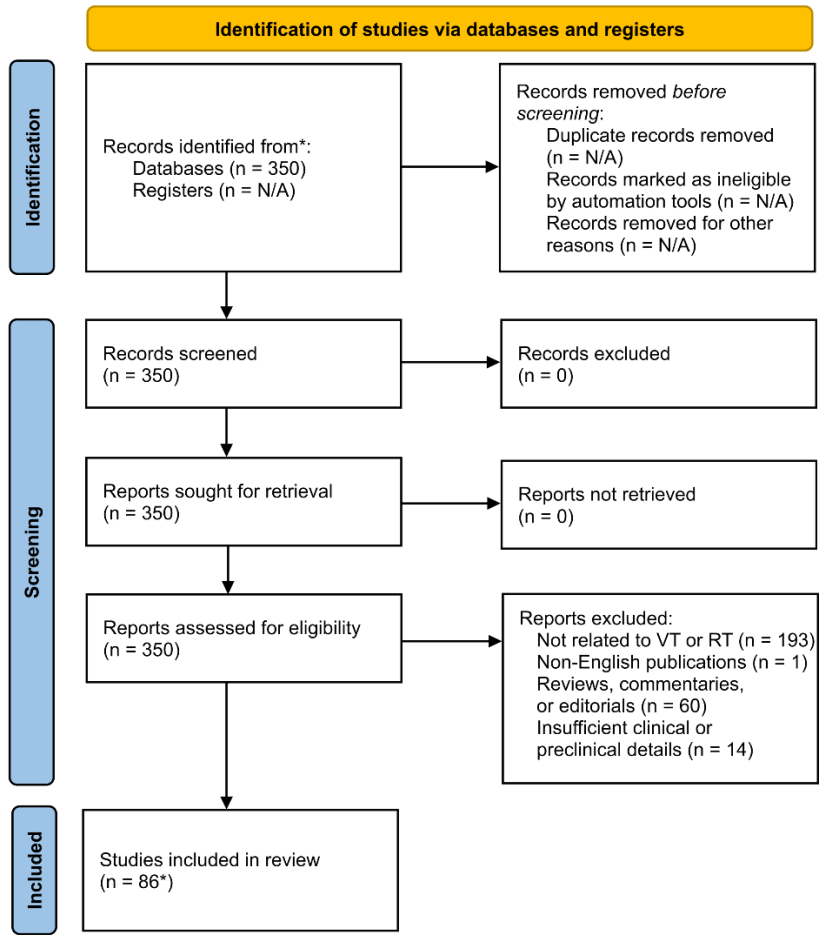

**Figure 1. PRISMA Flow Diagram of Study Selection Process.** Illustrates the systematic selection process for included studies, detailing the number of records identified, screened, excluded, and ultimately included in this review. Reasons for exclusion are categorized, including unrelated topics, insufficient data, duplicates, and non-English publications. (*4 studies were added after independent citation analysis).
Modified from Page et al [24].)

## 3. Results

### 3.1 Study Selection and Characteristics

A total of 350 studies were reviewed, and 86 studies were included in this review, comprising 12 preclinical studies, 49 case reports, 18 case series, and 7 clinical trials, published between 2015 and 2025. The PRISMA flow diagram summarizing the study selection process is presented in Figure 1.

Studies were conducted across 19 countries, with the majority originating from Europe (n = 45), followed by North America (n = 23), Asia (n = 15) and Australia (n = 1). A total of 349 patients were analyzed across the included studies, with 89% male patients. Nearly 90% of the studies utilized 6 MV photons as the primary treatment modality, with 25 Gy prescribed as the standard dose to the arrhythmic substrate. The methods for defining target volumes varied across studies. However, the majority of studies (n = 84) incorporated cardiac computed tomography (CT) for anatomical localization and 12-lead electrocardiography (ECG) for arrhythmic mapping. Advanced techniques such as EAM or electrocardiographic imaging (ECGI) were frequently employed (n = 78) to refine the target definition and delineate arrhythmogenic substrates.

Variability was observed in the reporting of outcomes, with some studies focusing on VT burden reduction and others prioritizing survival and toxicity as primary endpoints. Given the heterogeneity in study designs and follow-up durations, pooled analyses with subgroup considerations (e.g., treatment modalities and motion management strategies) were conducted to provide a comprehensive assessment of outcomes.

### 3.2 Preclinical Studies

A total of 12 preclinical studies explored the effects of STAR across diverse animal models, including pigs (n = 81), rabbits (n = 32), dogs (n = 25), and rats (n = 9), representing a combined total of 173 animals. These studies spanned five countries (USA, Japan, Germany, Russia, and South Korea) and provided critical insights into the efficacy and safety of STAR. The key characteristics and findings from these studies are summarized in Table 1. Studies assessed the impact of STAR on arrhythmia suppression, myocardial remodeling, conduction properties, and treatment safety. Photon-based STAR was the most commonly studied modality in preclinical studies (42%, n = 5), reflecting its widespread clinical adoption for arrhythmia management across various settings. Particle therapy modalities, including protons and carbon ions, were equally studied (33%, n = 4 each), highlighting growing interest in their precision for treating arrhythmogenic substrates. These studies assessed the impact of radiation on arrhythmia management, myocardial remodeling, and treatment safety. Amino et al. [25,26] demonstrated dose-dependent reductions in VT/VF inducibility, with higher doses leading to improved conduction parameters. Lehmann et al. [27,28] focused on achieving complete AV block with escalating doses, while Zei et al. [29] reported successful electrical isolation of the RSPV, emphasizing the impact of STAR on conduction pathways and arrhythmia suppression. Structural studies, such as those by Hohmann et al. [30,31] and Kancharla et al. [32], highlighted enhanced scar homogenization and stabilization of cardiac function post-MI. Molecular analysis from Kim et al. [33] provided insights into early proteomic changes linked to radiation-induced stress responses.

While STAR demonstrated promising efficacy, safety concerns were noted in specific studies. Takami et al. [34] reported pericardial effusion in irradiated rabbits, while Imamura et al. [35] observed conduction slowing and structural remodeling in long-term follow-up. Studies such as Vaskovskii et al. [36] explored photon therapy's impact on AV node ablation, demonstrating dose-dependent conduction block effects. These findings emphasize the need for precise dose optimization.

**Table 1.** Summary of Preclinical Studies Investigating Stereotactic Arrhythmia Radioablation (STAR) and Particle Therapy for Cardiac Applications.

| Authors (Year) | Country | Animal (n) | Disease Model | Modality (Dose Gy) | Key Findings |
|---|---|---|---|---|---|
| Lehmann et al (2015)[27] | USA/Germany | Pigs (4) | Explanted | Carbons (70/90/160 Gy) | No AV block up to 130 Gy; complete AV block at 160 Gy, confirmed by PET-CT; no visible myocardial damage. |
| Amino et al (2017) [25] | Japan | Dogs (8) | AF | Carbons (15 Gy) | VT/VF inducibility reduced (25% vs. 100%) (irradiated (n=4) vs non-irradiated (n=4)); improved conduction (QRS & RMS40); increased Cx43 (24-45%). |
| Lehmann et al (2017) [28] | USA | Pigs (10) | AV Junction Ablation | Photons (25/40/50/55 Gy) | Complete AV block achieved in 6/7 irradiated pigs (86%); lesion size increased with dose; no short-term side effects; no damage to esophagus, phrenic nerves, or trachea; histology revealed beam effects outside target volume. |
| Zei et al (2018) [29] | USA | Dogs + Pigs (19) | VT | Photons (15/20/25/35 Gy) | Successful electrical isolation of the RSPV achieved at 25 and 35 Gy (100%), partial isolation at 20 Gy (80%) and 15 Gy (50%); no complications or collateral tissue injury; transmural scar formation confirmed by histopathology. |
| Hohmann et al (2019) [30] | USA | Pigs (20) | LV Ablation (Healthy) | Protons (30/40 Gy) | Dose-dependent decline in LVEF (r = -0.69, P = .008); LV dilation correlated with dose (r = 0.75, P = .003); functional decline observed ~3 months post-treatment. |
| Hohmann et al (2020) [31] | USA | Pigs (14) | Post-MI | Protons (30/40 Gy) | Scar homogenization (treated: 30.1% myocytes vs. untreated: 59.9%); 4 VT-related sudden deaths; stable cardiac function; MRI revealed dose-related tissue effects over time. |

| Takami et al (2021) [34] | Japan | Rabbits (32) | Whole LV Irradiation | Carbons + Protons (25 Gy) | Significant LV conduction delays (PR: PT25 > control, P = .003); reduced P and QRS voltages; sustained effects at 6 months; VF induced in 1 carbon beam rabbit; no VF in proton group; mild-moderate pericardial effusion in 19% (carbon) and 44% (proton) with no tamponade. |
|---|---|---|---|---|---|
| Vaskovskii et al (2022) [36] | Russia | Pigs (2) | AV Node & LV Ablation | Photons (40/45 Gy) | 40 Gy induced transient AV block; 45 Gy resulted in permanent AV block and ventricular standstill by day 21; histology confirmed transmurality and precision. |
| Kim et al (2022) [33] | South Korea | Rats (9) | Proteomic (Healthy) | Photons (0/2/25 Gy) | 25 Gy induced significant proteomic changes within 7 days; early effects on signal transduction, adhesion, and stress response; upregulation of oxidative stress proteins; potential mediators of early anti-arrhythmic effects identified. |
| Amino et al (2023) [26] | Japan | Rabbits (26) | HC, AT/AF & VT/VF | Carbons (15 Gy) | Radiation reduced AT/AF (1.2% vs. 9.9%) and VT/VF (1.2% vs. 7.8%); improved conduction velocity; reversed Cx40/43 downregulation and sympathetic nerve sprouting. |
| Imamura et al (2023) [35] | USA | Pigs (19) | Normal + Infarcted Myocardium | Protons (40 Gy) | Reduced bipolar voltage amplitude (normal: 10.1→5.7 mV, infarcted: 2.0→0.8 mV); conduction velocity decreased (normal: 85→55 cm/s, infarcted: 43.7→26.3 cm/s); Cx43 reduction observed from 1-week post-irradiation; myocytolysis, capillary hyperplasia, and dilation at 8 weeks. |
| Kancharla et al (2024) [32] | USA | Pigs (10) | Post-MI VA | Photons (25 Gy) | SBRT reduced VA inducibility (100% vs. 25%, P=0.07); scar density increased (33% vs. 14%, P=0.07); no fibrosis in remote myocardium; SBRT improved scar homogenization. |

**Abbreviations:** AF: Atrial Fibrillation, AT: Atrial Tachycardia, AV: Atrioventricular, Cx40: Connexin-40, Cx43: Connexin-43, HC: Hypercholesterolemia, LVEF: Left Ventricular Ejection Fraction, MI: Myocardial Infarction, MRI: Magnetic Resonance Imaging, PET-CT: Positron Emission Tomography-Computed Tomography, PR: PR Interval, PT25: Proton Therapy 25 Gy, QRS: QRS Complex (ventricular depolarization), RMS40: Root Mean Square Voltage of the Last 40 ms, RSPV: Right Superior Pulmonary Vein, SBRT: Stereotactic Body Radiation Therapy, VT/VF: Ventricular Tachycardia/Fibrillation.

### 3.3 Case Reports

The 48 included case reports, published between 2015 and 2025, provided detailed insights into individual patient experiences with STAR for recurrent VT. These studies predominantly utilized photon-based STAR (n = 47), with 25 Gy in a single fraction being the standard dose prescription. Technologies used included LINAC systems, CyberKnife, and MRI-guided systems, with a variety of motion management strategies such as 4DCT, internal target volume (ITV) expansion, and fiducial marker-based tracking.

Most cases targeted monomorphic VT (MMVT) (n = 37), while polymorphic VT (PMVT) was less commonly reported (n = 12). PTV volumes varied significantly across cases, reflecting differences in arrhythmogenic substrate sizes and target delineation strategies. Notably, motion management techniques were adapted based on the technology used, with CyberKnife treatments employing fiducial markers and LINAC systems relying on 4DCT and ITV expansion.

Outcomes from these reports highlighted the efficacy of STAR in reducing VT burden (96.73% ± 6.41%), often achieving substantial suppression of arrhythmic episodes. While acute toxicities were rare, a few patients experienced pneumonitis or exacerbation of chronic obstructive pulmonary disease (COPD). However, long-term follow-up data were inconsistently reported, limiting the ability to draw definitive conclusions about the incidence and severity of late effects. A detailed summary of individual case reports, including treatment characteristics, is provided in Table 2.

**Table 2.** Summary of Case Reports Investigating Stereotactic Arrhythmia Radioablation (STAR) for Ventricular Tachycardia (VT).

| Author (Year) | Country | Age (Gender) | VT Type | Modality (Dose, fx) | Technology | PTV Volume (cc) | Motion Management |
|---|---|---|---|---|---|---|---|
| Loo et al (2015) [18] | USA | 71 (M) | MMVT | Photons (25 Gy, 1fx) | CyberKnife | NR | Fiducial Marker |
| Jameau et al (2018) [37] | Switzerland | 75 (M) | PMVT | Photons (25 Gy, 1fx) | CyberKnife | 21 | Fiducial Marker |
| Haskova et al (2018) [38,39] | Czech Republic | 34 | PMVT | Photons (25 Gy, 1fx) | CyberKnife | 62.2 | NR |
| Bhaskaran et al (2019) [40] | Canada | 34 (F) | MMVT | Photons (25 Gy, 1fx) | LINAC | 52 | 4DCT, ITV |
| Zeng et al (2019) [41] | China | 29 (M) | PMVT | Photons (25 Gy, 1fx) | CyberKnife | 71.22 | NR |
| Marti'-Almor et al (2020) [42] | Spain | 64 (M) | MMVT | Photons (25 Gy, 1fx) | LINAC | NR | 4DCT |

| Study | Country | Age (Sex) | Type | Radiation (Dose) | Device | Volume | Motion Management |
|---|---|---|---|---|---|---|---|
| Narducci et al (2020) [43] | Italy | 60 (M) | MMVT | Photons (25 Gy, 1fx) | LINAC | 303 | 4DCT, ITV |
| Mayinger et al (2020) [44] | Switzerland | 71 (M) | MMVT | Photons (25 Gy, 1fx) | MRIdian | 115.1 | NR |
| Krug et al (2020) [45] | Germany | 78 (M) | MMVT | Photons (25 Gy, 1fx) | LINAC | 42.2 | NR |
| Park and Choi (2020) [46] | South Korea | 76 (M) | MMVT | Photons (25 Gy, 1fx) | LINAC | NR | NR |
| Dusi et al (2021) [47] | Italy | 73 (M) | MMVT | Photons (25 Gy, 1fx) | NR | 27.7 | NR |
| Peichl et al (2021) [48] [39] | Czech Republic | 66 (M) | MMVT | Photons (25 Gy, 1fx) | CyberKnife | 18.3 | NR |
| Amino et al (2021) [49] | Japan | 75 (F) | PMVT | Photons (25 Gy, 1fx) | LINAC | 49.7 | NR |
| Quick et al (2021) [50] | Germany | 85 (M) | MMVT | Photons (25 Gy, 1fx) | NR | 8.51, 15.01 | NR |
| Lee et al (2021) [51] | Korea | 11 (M) | MMVT | Photons (25 Gy, 1fx) | LINAC | NR | 4DCT, ITV |
| Kautzner et al (2021) [52] | Czech Republic | 52 (M) | MMVT | Photons (25 Gy, 1fx) | LINAC | 52 | NR |
| | | 57 (M) | MMVT | Photons (25 Gy, 1fx) | LINAC | 62.1 | NR |
| | | 67 (M) | PMVT | Photons (25 Gy, 1fx) | LINAC | 70 | NR |
| Thosani et al (2021) [53] | USA | 73 (M) | MMVT | Photons (25 Gy, 1fx) | LINAC | 62.6 | Margin |
| Aras et al (2021) [54] | Turkey | 58 (M) | MMVT | Photons (25 Gy, 1fx) | LINAC | NR | 4DCT, ITV |
| Li et al (2022) [55] | China | 54 (M) | MMVT | Photons (25 Gy, 1fx) | LINAC | 74.7 | 4DCT, ITV |
| Hayase et al (2022) [56] | USA | 78 (M) | MMVT | Photons (25 Gy, 1fx) | LINAC | NR | NR |
| Levis et al (2022) [57] | Italy | 73 (M) | MMVT | Photons (25 Gy, 1fx) | LINAC | 89 | 4DCT |
| Haskova et al (2022) [39] | Czech Republic | 77 (M) | NR | Photons (25 Gy, 1fx) | CyberKnife | 14.3 | NR |
| Huang et al (2022) [58] | Taiwan | 63 (M) | MMVT | Photons (12 Gy, 1fx) | LINAC | 65.75 | ITV |
| van der Ree et al (2022) [59] | Netherlands | 60 (M) | PMVT | Photons (25 Gy, 1fx) | LINAC | 300 | 4DCT, ITV |
| Wutzler et al (2022) [60] | Germany | 56 (M) | PMVT | Photons (25 Gy, 1fx) | LINAC | NR | 4DCT, ITV |

| Study | Country | Age (Sex) | Technique | Modality | Machine | Volume | Motion Management |
|---|---|---|---|---|---|---|---|
| Bernstein et al (2022) [61] | USA | 75 (M) | MMVT | Photons (25 Gy, 1fx) | LINAC | 87.9 | 4DCT |
| Kurzelowski et al (2022) [62] | Poland | 69 (M) | MMVT | Photons (25 Gy, 1fx) | LINAC | 56.37 | DIBH |
| | | 72 | MMVT | Photons (25 Gy, 1fx) | LINAC | 56.72 | DIBH |
| Cybulska et al (2022) [63] | Poland | 67 (M) | PMVT | Photons (25 Gy, 1fx) | LINAC | NR | DIBH |
| Ninni et al (2022) [64] | France | 42 (M) | MMVT | Photons (25 Gy, 1fx) | CyberKnife | NR | NR |
| Nasu et al (2022) [65] | Japan | 58 (M) | PMVT | Photons (25 Gy, 1fx) | LINAC | 29.1 | 4DCT |
| Pavone et al (2022) [66] | Italy | 73 (M) | PMVT | Photons (25 Gy, 1fx) | LINAC | NR | 4DCT, ITV |
| Cozzi et al (2022) [67] | Italy | 81 (M) | MMVT | Photons (25 Gy, 1fx) | LINAC | 122.5 | 4DCT, ITV |
| Mehrhof et al (2023) [68] | Germany | 54 (M) | MMVT | Photons (25 Gy, 1fx) | CyberKnife | 75.2 | Fiducial Marker |
| | | 61 (M) | PMVT | Photons (25 Gy, 1fx) | LINAC | 134.6 | ITV |
| Jiwani et al (2023) [69] | USA | 83 (M) | MMVT | Photons (25 Gy, 1fx) | LINAC | 146.7 | 4DCT |
| van der Ree et al (2023) [70] | Netherlands | 47 (F) | PMVT | Photons (2 Gy, 2fx; 20 Gy, 1fx) | CyberKnife | 16 | Fiducial Tracking |
| Kaestner et al (2023) [71] | Germany | 63 (F) | MMVT | Photons (25 Gy, 1fx) | LINAC | NR | NR |
| Wijesuriya et al (2023) [72] | UK | 69 (F) | MMVT | Photons (25 Gy, 1fx) | LINAC | NR | NR |
| Vaskovskii et al (2023) [73] | Russia | 57 (M) | MMVT | Photons (25 Gy, 1fx) | LINAC | 46 | 4DCT, ITV |
| Vozzolo et al (2023) | USA | 44 (M) | MMVT | Photons (25 Gy, 1fx) | LINAC | NR | 4DCT |
| Keyt et al (2023) [74] | USA | 75 (M) | MMVT | Photons (25 Gy, 1fx) | LINAC | 85 | 4DCT, ITV |
| Amino et al (2024) [75] | Japan | 60 (M) | MMVT | Carbons (25 Gy, 1fx) | XiO (Elekta) | 29.7 | Motion Margin |
| Kautzner et al (2024) [76] | Czech Republic | 54 (F) | MMVT | Photons (25 Gy, 1fx) | CyberKnife | NR | NR |
| Kaya et al (2024) [77] | Netherlands | 72 (M) | MMVT | Photons (25 Gy, 1fx) | LINAC | 11* | 4DCT |
| Trinh et al (2025) [78] | USA | 62 (M) | MMVT | Photons (25 Gy, 1fx) | LINAC | NR | NR |

**Abbreviations:** 4DCT: Four-Dimensional Computed Tomography, DIBH: Deep Inspiration Breath Hold, ITV: Internal Target Volume, LINAC: Linear Accelerator, MMVT: Monomorphic Ventricular Tachycardia, NR: Not Reported, PMVT: Polymorphic Ventricular Tachycardia. **Note:** * indicates Clinical Target Volume (CTV)

## 3.4 Clinical Series: Case Series and Clinical Trials

A total of 24 clinical trials and case series were included, with sample sizes ranging from 3 to 36 patients. These studies provided valuable insights into the efficacy of STAR for recurrent ventricular tachycardia (VT) in larger cohorts. Across these studies, PTV values varied significantly, with a median of 81.55 cc (range: 14-372 cc), reflecting variability in target delineation practices and arrhythmogenic substrate sizes. Margins used for target volume expansion were inconsistent, ranging from 1 mm to 8 mm isotropic expansion, further emphasizing the variability in contouring practices across institutions.

Most studies employed photon-based STAR, with doses predominantly prescribed at 25 Gy in a single fraction (n = 22, 91.7%). Of these, LINAC-based systems were used in 79.2% of cases (n = 19), while CyberKnife treatments were reported in 16.7% (n = 4). One study used MRI-guided STAR system, showcasing emerging technology for arrhythmia ablation.

Patient characteristics revealed significant baseline cardiac dysfunction, with a median LVEF of 27.0% (range: 10-72%). Ischemic cardiomyopathy (ICM) and non-ischemic cardiomyopathy (NICM) were nearly equally distributed, representing 51.81% and 48.19% of patients, respectively. Motion management techniques included 4DCT, ITV expansion, and fiducial marker-based tracking. 4DCT was the most frequently employed strategy (62.5%), particularly in LINAC-based treatments, while fiducial tracking was utilized for CyberKnife treatments. However, details on motion management were inconsistently reported in some studies, limiting the ability to evaluate specific trends.

On average, a 75.0% reduction in VT burden was observed at six months, highlighting the substantial arrhythmic suppression achieved with STAR. Details of the included clinical trials and case series, including patient demographics, cardiomyopathy classification, and treatment characteristics, are summarized in Table 3.

**Table 3.** Summary of Clinical Trials and Case Series Investigating Stereotactic Arrhythmia Radioablation (STAR) for Ventricular Tachycardia (VT). If the entry is a clinical trial, its trial name is reported in the Author column.

| Author (Year) | Country | Sample Size | Age (Median, | Gender (M/F) | CM | LVEF(%) (Median, | PTV Volume | Modality (Dose, fx) | Technology |
|---|---|---|---|---|---|---|---|---|---|

| Study | Country | N | Age (Median, Range) | Sex | Cardiomyopathy | LVEF (Median, Range) | Target Volume (cc) | Radiation (Dose, fx) | Device |
|---|---|---|---|---|---|---|---|---|---|
| Cuculich et al (2017) [19] | USA | 5 | 62 (60-83) | 4M/1F | 2 ICM; 3 NICM | 22 (15-26) | 51.3 (17.3-81) | Photons (25 Gy, 1fx) | LINAC |
| Robinson et al (2019) (ENCORE-VT) [79] | USA | 19 | 66 (49-81) | 17M/2F | 11 ICM; 8 NICM | 25 (15-58) | 98.9 (60.9-298.8) | Photons (25 Gy, 1fx) | LINAC |
| Chin et al (2020) [80] | USA | 8 | 74 (65-86) | 8M | 4 ICM; 4 NICM | 20 (15-32.5) | 84.9 (21.1-190.7) | Photons (15-25 Gy, 1fx) | LINAC |
| Gianni et al (2020) [81] | USA | 5 | 67 (45-76) | 5M | 4 ICM; 1 NICM | 25 (20-55) | 173 (80-184) | Photons (25 Gy, 1fx) | CyberKnife |
| Lee et al (2021) [82] | UK | 7 | 70 (60-79) | 4M/3F | 5 ICM; 2 NICM | 25 (15-45) | 89.5 (57.5-139) | Photons (25 Gy, 1fx) | LINAC |
| Yugo et al (2021) [83] | Taiwan | 3 | 68 (65-83) | 2M/1F | 3 NICM | 44 (20-59) | 70 (20-130) | Photons (25 Gy, 1fx) | LINAC |
| Ho et al (2021) [84] | USA | 6 | 72.5 (64-77) | 6M | 2 ICM; 4 NICM | 26 (10-46) | 120.5 (66-193) | Photons (25 Gy, 1fx) | LINAC |
| Carbucicchio et al (2021) (STAR-MI-VT) [85] | Italy | 7 | 72 (59-78) | 7M | 3 ICM; 4 NICM | 21.1 (20.3-44.4) | 198.3 (88.1-239) | Photons (25 Gy, 1fx) | LINAC |
| Qian et al (2022) [86] | USA | 6 | 72 (70-73) | 6M | 6 ICM | 20 (16-20) | 319 (280-330) | Photons (25 Gy, 1fx) | LINAC |
| Wight et al (2022) [87,88] | USA | 14 | 60.5 (50-70) | 10M/4F | 5 ICM; 9 NICM | NR | NR | Photons (25 Gy, 1fx) | LINAC |
| Molon et al (2022) [89] | Italy | 6 | 79.5 (61-85) | 5M/1F | 3 ICM; 3 NICM | 26.5 (20-42) | NR | Photons (25 Gy, 1fx) | LINAC |
| Ninni et al (2022) [90] | France | 17 | 68 (30-83) | 13M/4F | 10 ICM; 7 NICM | 35 (20-53) | 53.3 (19.96-185.88) | Photons (25 Gy, 1fx) | CyberKnife |
| Chang et al (2023) [91] | Korea | 6 | 72 (63-85) | 4M/2F | 3 ICM; 3 NICM | 31.5 (24-57) | 52.2 (17.5-246.8) | Photons (25 Gy, 1fx) | LINAC |
| Aras et al (2023) [92] | Turkey | 8 | 61.5 (33-85) | 8M | 2 ICM; 6 NICM | 25 (10-30) | 157.4 (70.5-272.7) | Photons (25 Gy, 1fx) | LINAC |
| van der Ree et al (2023) | Netherlands | 6 | 73 (54-83) | 6M | 6 ICM | 38 (24-52) | 187 (93-372) | Photons (25 Gy, 1fx) | LINAC |

| Study | Country | N | Age | Sex | CM type | PTV (cc) | Follow-up (mo) | Modality | System |
|---|---|---|---|---|---|---|---|---|---|
| STARNL-1 [93] | | | | | | | | | |
| Krug et al (2023) RAVENTA [94] | Germany | 5 | 67 (49-74) | 4M/1F | 2 ICM; 3 NICM | 35 (20-45) | 69.6 (43.4-80.7) | Photons (25 Gy, 1fx) | NR |
| Herrera Siklody et al (2023) [95] | Switzerland | 20 | 68 (47-80) | 15M/5F | 6 ICM; 14 NICM | 31 (20-72) | 23 (14-115) | Photons (20-25 Gy, 1fx) | CyberKnife/ MRIdian/ LINAC |
| Miszczyk et al (2023) SMART-VT [96] | Czech Republic | 11 | 67 (45-72) | 10M/1F | 9 ICM; 2 NICM | 27 (20-40) | 73 (18.6-111.3) | Photons (25 Gy, 1fx) | LINAC |
| Amino et al (2023) (SRAT) [97] | Japan | 3 | 71 (60-91) | 1M/2F | 1 CIM; 2 NICM | 27 (20-65) | 55 (49.7-96.4) | Photons (25 Gy, 1fx) | LINAC |
| Haskova et al (2024) [98] | Czech Republic | 36 | 66 (56-76) | 33M/3F | 20 ICM; 16 NICM | 31 (22,40) | 39.4 (12.6-90.5) | Photons (25 Gy, 1fx) | CyberKnife |
| Arkles et al (2024) [99] | USA | 15 | 65 (57.2-72.8) | 13M/2F | 7 ICM; 8 NICM | 30.2 (26.6-33.8) | 45.6 (84.7-124.1) | Photons (25 Gy, 1fx) | LINAC |
| Borzov et al (2024) [100] | Israel | 3 | 64 (63-72) | 3M | 1 ICM; 2 NICM | 27.5 (15-30) | 49.7 (47.8-91.8) | Photons (25 Gy, 1fx) | LINAC |
| Bianchi et al (2024) [101] | Italy | 11 | 68 (53-81) | 11M | 5 ICM; 6 NICM | 40 (30-57) | 90.4 (30.6-119.5) | Photons (25 Gy, 1fx) | MRIdian |
| Das et al (2025) [102] | Australia | 12 | 74.9 (63.5-86.1) | 10M/2F | 9 ICM; 3 NICM | 20 (15.3-31.5) | 135.1 (27.4-226.5) | Photons (25 Gy, 1fx) | LINAC |

**Abbreviations:** CM: Cardiomyopathy, CyberKnife: Robotic Radiosurgery System, ICM: Ischemic Cardiomyopathy, LINAC: Linear Accelerator, LVEF: Left Ventricular Ejection Fraction, MRIdian: MRI-Guided Radiation Therapy System, NICM: Non-Ischemic Cardiomyopathy, NR: Not Reported, PTV: Planning Target Volume, fx: Fraction(s).

### 3.5 Pooled-Analysis

3.5.1 6-Month and 12-Month Mortality

The pooled proportion of deaths at 6 months was 15.5% (95% CI: 11-20%), with minimal heterogeneity across studies ($I^2$ = 0.00%, Cochran's Q = 18.6, p = 0.67). For 12-month mortality, the pooled estimate was 32.5% (95% CI: 26.7-38.3%), also demonstrating low heterogeneity ($I^2$ = 0.00%, Cochran's Q = 16.47, p = 0.63). Figures 2(a) and 2(b) present the forest plots for 6-month and 12-month mortality, respectively.

### 3.5.2 Grade 3+ Acute Toxicities

The pooled rate of grade 3+ adverse events within 90 days of treatment was 7.2% (95% CI: 4.2-10.3%), with no observed heterogeneity ($I^2$ = 0.00%, Cochran's Q = 7.6, p = 0.99). Toxicities included heart failure (n=3), and esophagitis (n = 2). The forest plot for acute toxicity rates is shown in Figure 2(c).

### 3.5.3 VT Events Reduction at 6 Months

The pooled percentage reduction in VT events at 6 months was 75.4% (95% CI: 73.4-77.4%), with substantial heterogeneity ($I^2$ = 98.80%, Cochran's Q = 1328.4, $p < 0.05$). Figure 2(d) illustrates the forest plot for VT event reduction.

### 3.5.4 Subgroup Analysis

Subgroup analyses provided additional insights into factors influencing outcomes. Comparisons between LINAC- and CyberKnife-based treatments revealed similar mortality rates at 12 months (35%, 95% CI: 27-38% for LINAC vs. 29%, 95% CI: 20-39% for CyberKnife), with minimal differences in acute toxicity (8%, 95% CI: 4-13% for LINAC vs. 6%, 95% CI: 1-12% for CyberKnife). Age-stratified analysis showed slightly lower mortality at 6 months for patients younger than the median age (14%, 95% CI: 7-21%) compared to older patients (19%, 95% CI: 11-28%). Similarly, patients with LVEF above the median had marginally lower mortality at 6 months (13%, 95% CI: 7-19%) compared to those with LVEF below the median (20%, 95% CI: 13-27%). Regarding cardiomyopathy types, mortality at 6 months was similar for patients with NICM (16%, 95% CI: 9-22%) compared to ICM (14%, 95% CI: 7-21%), while VT burden reduction was higher for NICM group (99% for NICM vs. 59% for ICM). In contrast, VT burden reduction at 6 months demonstrated substantial variation across subgroups. For example, NICM patients showed markedly higher VT reduction (99%) compared to ICM patients (59%), while patients with higher LVEF or younger age also showed greater reductions. These subgroup findings are visualized in Figure 3, with the full summary provided in Table 4. Additional forest plots for mortality and acute toxicity subgroup analyses are included in the Supplement (Figures S1–S3).

**Table 4**. Summary of Pooled-Analysis Results for Stereotactic Arrhythmia Radioablation (STAR) in Ventricular Tachycardia (VT): Outcomes include pooled estimates for mortality at 6 and 12 months, reduction in VT burden at 6 months, and grade 3+ adverse events within 90 days. Subgroup analyses evaluate variations by treatment modality (LINAC vs. CyberKnife), LVEF (≤ median vs. > median), patient age (≤ median vs. > median), and cardiomyopathy type

(ICM vs. NICM). Results are presented as pooled effect estimates with 95% confidence intervals (CI), Cochran's Q statistics, and heterogeneity (I²).

| Outcome | Metric | Overall | LINAC vs CyberKnife | LVEF (≤ Median vs > Median) | Age (≤ Median vs > Median) | Cardiomyopathy (ICM vs NICM) |
|---|---|---|---|---|---|---|
| Deaths at 6 months | Pooled Effect | 0.16 (0.11, 0.20) | 0.16 (0.11,0.22) vs 0.12 (0.02,0.21) | 0.20 (0.13,0.27) vs 0.13 (0.07,0.19) | 0.14 (0.09,0.19) vs 0.19 (0.11,0.28) | 0.14 (0.07,0.21) vs 0.16 (0.09,0.22) |
| | Cochran's Q | 18.6 (p=0.67) | 17.7 (p=0.54) vs 0.16 (p=0.92) | 6.44 (p=0.89) vs 9.96 (p=0.35) | 10.72 (p=0.55) vs 6.84 (p=0.65) | 3.3 (p=0.86) vs 11.6 (p=0.31) |
| | I² | 0.00% | 0.00% vs 0.00% | 0.00% vs 9.6% | 0.00% vs 0.00% | 0.0% vs 13.6% |
| Deaths at 12 months | Pooled Effect | 0.33 (0.27, 0.38) | 0.35 (0.26,0.42) vs 0.29 (0.20,0.39) | 0.35 (0.26,0.44) vs 0.31 (0.23,0.38) | 0.31 (0.24,0.37) vs 0.38 (0.27,0.50) | 0.34 (0.26,0.42) vs 0.30 (0.21,0.40) |
| | Cochran's Q | 16.47 (p=0.63) | 6.94 (p=0.96) vs 8.71 (p=0.03) | 4.54 (p=0.92) vs 10.95 (p=0.18) | 13.52 (p=0.26) vs 1.79 (p=0.97) | 8.29 (p=0.41) vs 6.93 (p=0.44) |
| | I² | 0.00% | 0.00% vs 65.57% | 0.00% vs 29.8% | 18.64% vs 0.00% | 3.48% vs 0.00% |
| Grade 3+ Adverse Events within 90 days | Pooled Effect | 0.07 (0.04, 0.10) | 0.07 (0.03,0.11) vs 0.06 (0.01,0.12) | 0.07 (0.03,0.12) vs 0.06 (0.02,0.10) | 0.06 (0.01,0.12) vs 0.07 (0.03,0.11) | 0.07 (0.02,0.11) vs 0.07 (0.02,0.11) |
| | Cochran's Q | 7.62 (p=0.99) | 5.51 (p=0.99) vs 2.07 (p=0.56) | 2.33 (p=0.99) vs 5.18 (p=0.88) | 1.35 (p=0.99) vs 6.23 (p=0.93) | 3.36 (p=0.91) vs 3.32 (p=0.98) |
| | I² | 0.00% | 0.00% vs 0.00% | 0.00% vs 0.00% | 0.00% vs 0.00% | 0.00% vs 0.00% |
| VT events reduction at 6 months | Pooled Effect | 0.75 (0.73, 0.77) | 0.98 (0.97,1.00) vs 0.15 (0.02,0.28) | 0.69 (0.66,0.72) vs 0.89 (0.79,0.99) | 0.77 (0.74,0.80) vs 0.83 (0.73,0.93) | 0.59 (0.54,0.64) vs 0.99 (0.97,1.01) |
| | Cochran's Q | 1328.4 (p<0.05) | 62.24 (p<0.05) vs 26.42 (p<0.05) | 742.60 (p<0.05) vs 5.11 (p=0.40) | 760.51 (p<0.05) vs 7.38 (p=0.29) | 405.52 (p<0.05) vs 13.38 (p=0.06) |

| | | | | | |
|---|---|---|---|---|---|
| | I² | 98.80% | 77.51% vs 96.21% | 98.65% vs 2.12% | 98.82% vs 18.68% | 98.52% vs 47.69% |

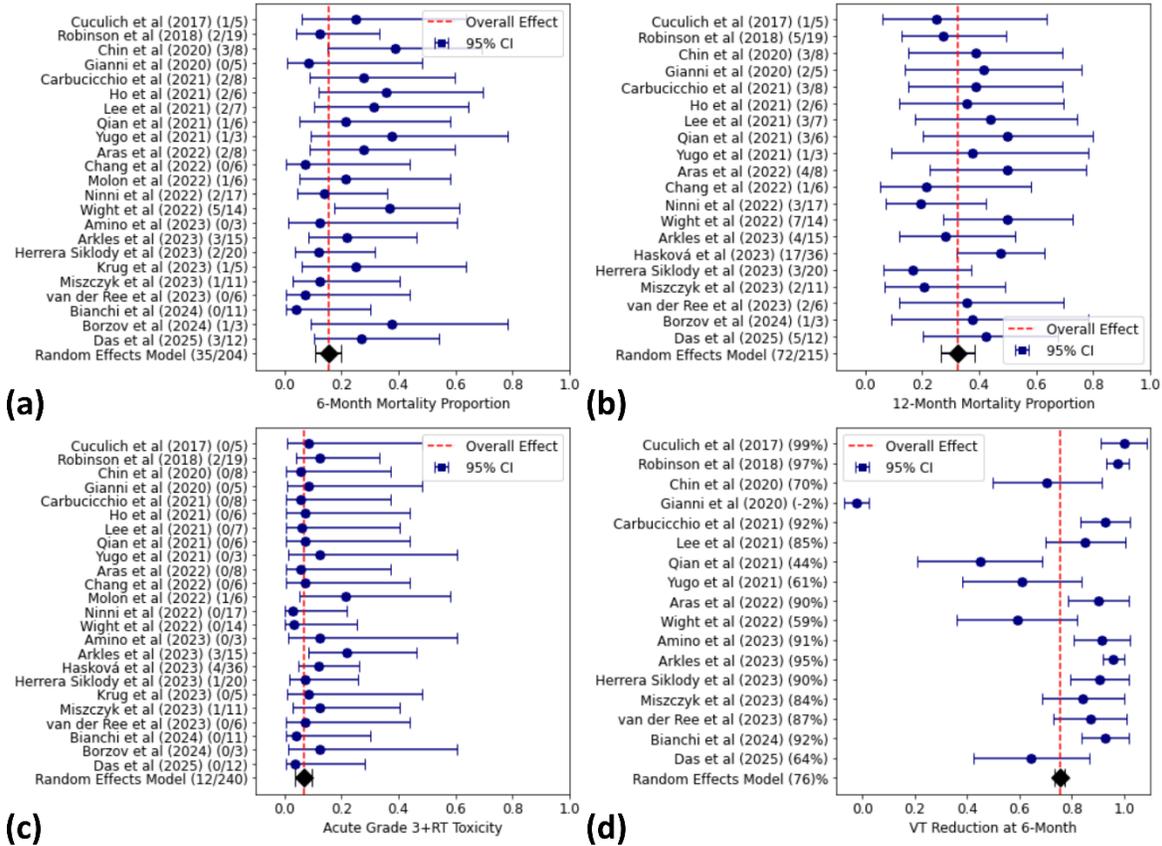

**Figure 2. Forest Plots Summarizing Pooled-Analysis Results for Mortality, VT Burden Reduction, and Acute Toxicity.** Forest plots displaying the pooled effect estimates and 95% confidence intervals (CI) for **(a)** 6-month mortality, **(b)** 12-month mortality, **(c)** acute grade 3+ toxicity rates within 90 days, and **(d)** VT reduction at 6 months. The red dashed line represents the overall effect, while individual blue points and bars represent study-specific estimates and their CIs. A random-effects model was used for pooled-analysis.

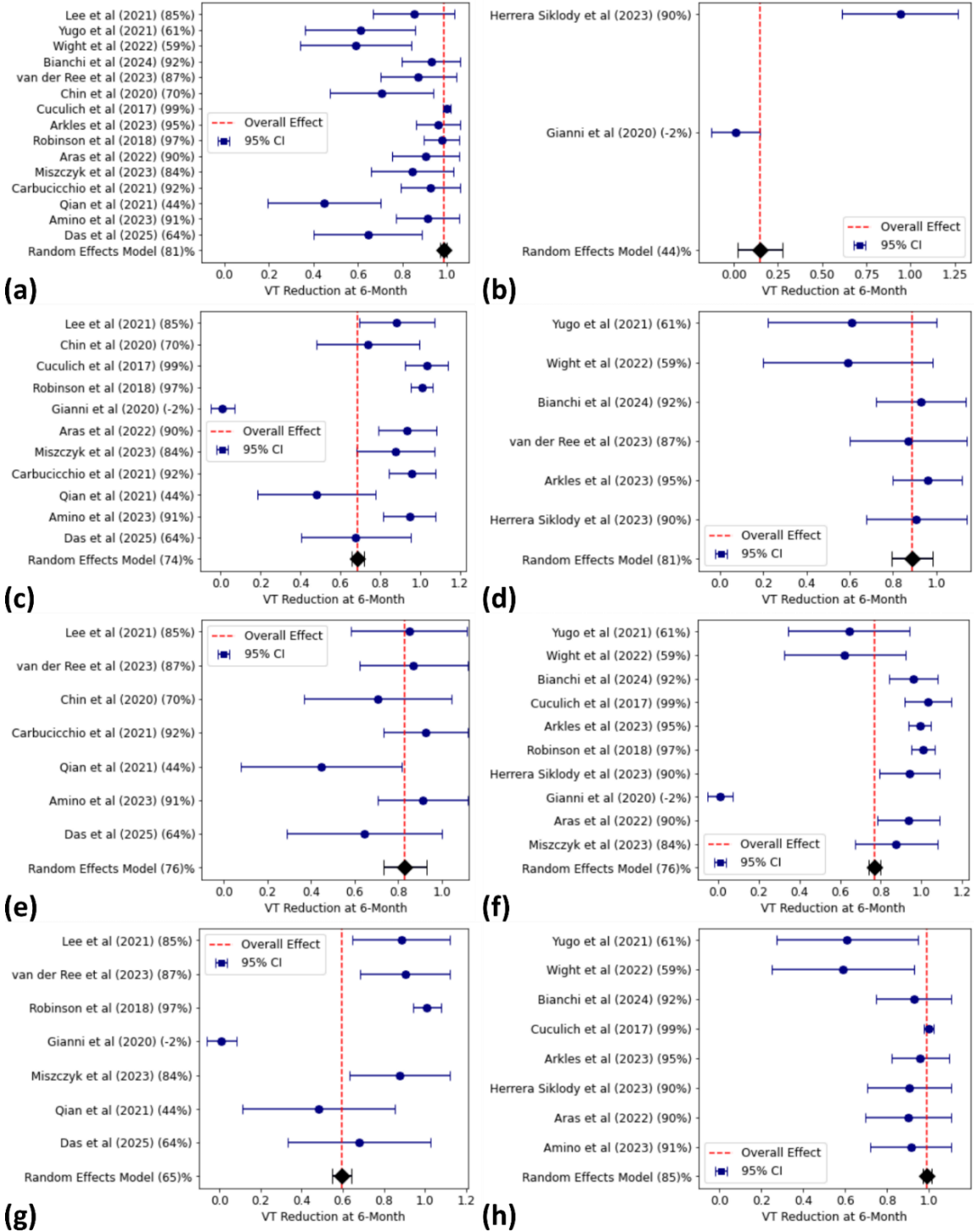

**Figure 3.** Forest plots depicting pooled VT burden reduction at 6 months following stereotactic arrhythmia radioablation (STAR), stratified by key subgroups. Each plot shows individual study estimates with 95% confidence intervals and an overall pooled estimate using a random-effects model. Studies are labeled with their reported VT reduction (%) and ordered chronologically. Subgroup comparisons include: **(A–B)** treatment modality — LINAC-based STAR **(A)** vs. CyberKnife **(B)**, **(C–D)** baseline left ventricular ejection fraction (LVEF) ≤ median **(C)** vs. > median **(D)**, **(E–F)** patient age ≤ median **(E)** vs. > median **(F)**, and **(G–H)** underlying cardiomyopathy — ischemic (ICM) **(G)** vs. non-ischemic (NICM) **(H)**.

## 4. Discussion

This narrative review and exploratory pooled analysis provides an updated synthesis of STAR for refractory ventricular tachycardia, integrating findings across preclinical reports, case series, and prospective clinical trials. The data reinforce that STAR is primarily utilized in a high-risk, heavily pretreated patient population—many of whom have failed conventional therapies such as catheter ablation and antiarrhythmic drugs. Despite this clinical complexity, early and intermediate outcomes appear promising, particularly with respect to sustained reductions in VT burden and low rates of acute severe toxicity. However, variability in follow-up intervals, reporting standards, and endpoint definitions limits direct comparability across studies, underscoring the need for harmonized prospective protocols.

Subgroup analyses revealed important insights into factors influencing outcomes. Younger patients and those with higher LVEF consistently demonstrated better survival and VT reduction rates. As compared to patients with ICM, NICM patients experienced superior VT burden reduction (99% vs. 59%) and reduced mortality at 12 months (34% vs 30%). These findings suggest that STAR outcomes may vary based on patient-specific characteristics, highlighting the need for tailored treatment approaches and stratified clinical trial designs.

The high heterogeneity observed in VT burden reduction outcomes ($I^2$ = 98.80%) highlights a critical need for standardized reporting and consistent methodologies. Definitions of VT burden varied across studies, with some quantifying episodes per unit time and others measuring total VT events. Additionally, while our pooled analysis standardized the evaluation at a 6-month timepoint, the definitions of pre- and post-treatment intervals and blanking periods were not uniform across studies. Some studies applied a 6-week blanking period to exclude early arrhythmic events post-treatment, while others did not mention such adjustments. Moreover, VT burden metrics varied, including sustained VT, all VT/VF events, or ICD interventions. These inconsistencies highlight the need for standardized definitions and outcome intervals in future STAR reporting. Such variability complicates comparisons and pooled-analyses, emphasizing the importance of establishing reporting frameworks akin to TRIPOD [103] standards for AI studies. Additionally, individual studies sometimes contributed disproportionately to pooled estimates. For example, one study (Haskova et al [98]) reported 17 of 36 deaths at 12 months but did not provide short-term outcomes, potentially skewing the 12-month mortality estimate. These imbalances further underscore the importance of reporting outcomes at standardized time intervals to improve cross-study comparability.

Subgroup analyses identified younger patients, NICM, and higher LVEF as predictors of favorable outcomes. Future clinical trials should stratify patients based on these factors and consider incorporating interim analyses or predefined endpoints, such as VT-free survival, to evaluate efficacy or safety. Trials may also include provisions for early conclusion if predefined thresholds for success or excessive adverse events are met, ensuring patient safety and resource optimization. This approach would not only improve trial efficiency but also minimize risk for high-risk patients. The results also highlight the need to explore differential outcomes between LINAC- and CyberKnife-based treatments, particularly in terms of toxicity profiles and cost-effectiveness. Accurate motion management remains a cornerstone of STAR. LINAC-based systems predominantly rely on 4DCT and ITV expansions, while CyberKnife employs fiducial tracking to accommodate respiratory and cardiac motion. Although CyberKnife offers sub-millimeter precision, its treatment times are significantly longer compared to LINACs (120 minutes vs 30 minutes), posing logistical challenges in a clinical setting. The variability in PTV margins across studies (ranging from 1 mm to 8 mm isotropic expansions) further underscores the lack of standardization in STAR planning. Addressing these inconsistencies is critical for optimizing treatment precision and minimizing radiation dose to surrounding organs.

Particle therapies, such as protons and carbon ions, represent an emerging frontier in STAR, particularly for younger patients or those with complex anatomies. The ability to leverage the Bragg peak for precise dose deposition makes these modalities uniquely suited for cases involving critical adjacent structures, such as the esophagus and lungs. Preclinical studies have demonstrated the feasibility of particle therapy for arrhythmia ablation [34,35]; however, clinical data remain sparse. Dusi et al [47] and Amino et al [75] successfully demonstrated the first-in-human use of protons and carbons to treat VT and demonstrated a reduction in VT events post STAR.

Lee et al [51] treated an 11-year-old pediatric patient for VT with photons. While the patient was in good condition at the 3-month follow-up visit, long-term follow-up data are unavailable, and this patient might have benefited from proton therapy, given its dosimetric advantages. Shah et al [104] demonstrated significant reductions in OAR doses for retrospectively planned patients treated with proton therapy compared with corresponding photon plans. Despite these promising developments, challenges such as range uncertainties, motion management, and the high costs of particle therapy must be addressed to facilitate its broader adoption.

Grade 3+ adverse events were rare (7%, 95% CI: 4-10%) but underscore the importance of meticulous planning to avoid significant complications. The reported cases of esophagitis and pneumonitis, particularly in patients with posterior substrates, emphasize the need for advanced planning techniques and possibly proton therapy to spare critical structures. Moreover, attention must be paid to the radiation dose delivered to implantable cardioverter-defibrillators (ICDs), as inappropriate shocks or device malfunctions remain a concern. Guidelines such as AAPM TG-203 [105] offer practical recommendations for managing these challenges during treatment.

Cases requiring repeated STAR treatments illustrate the challenges of achieving complete arrhythmic suppression in patients with extensive or complex arrhythmogenic substrates. Improved target identification, supported by EAM, ECGI, and advanced imaging modalities, can mitigate the need for retreatments. Artificial intelligence-based automated models and semi-automated have shown promise in automating substrate delineation, reducing inter-observer variability, and identifying non-responders earlier in the treatment course [106–109].

The efficacy of STAR is underpinned by its ability to induce fibrosis and alter myocardial conduction properties, disrupting arrhythmogenic circuits. However, the biological mechanisms remain incompletely understood. Preclinical studies have highlighted changes in gap junction remodeling (e.g., connexin-43 expression), conduction slowing, and fibrosis as key contributors to arrhythmia suppression [25,26]. Long-term consequences of radiation, including vascular damage and inflammatory responses, require further investigation, particularly in younger patients who may face increased risks of late toxicities.

The overwhelming majority of STAR studies have been conducted in North America, Europe, and East Asia, with limited representation from South America, Africa, the Middle East, and the Indian subcontinent. This geographic disparity reflects broader inequities in access to advanced radiation therapy technologies. Efforts must be made to globalize radiation therapy, making STAR accessible to all patients who need it. This includes reducing financial and logistical barriers to acquiring treatment infrastructure and fostering international collaboration to ensure equitable access.

This study has several limitations. First, although the pooled analyses were conducted using established meta-analytic techniques, the review itself was not prospectively registered (e.g., on PROSPERO), and a formal risk-of-bias assessment was not performed. This is primarily because the review was designed as a narrative synthesis with exploratory pooled analysis rather than a formal systematic review. We focused on hypothesis generation using study-level data, acknowledging the variability in study designs, reporting quality, and follow-up durations. Second, the use of a single bibliographic database (PubMed) may limit search comprehensiveness, though we mitigated this by manually reviewing reference lists and citations of included studies. Third, high heterogeneity was observed—particularly in VT burden reduction outcomes—reflecting inconsistency in endpoint definitions, blanking periods, and reporting standards. Fourth, unmeasured confounding may have influenced the results. Variables such as antiarrhythmic drug use, NYHA status, and comorbidities were inconsistently reported, limiting adjustment in pooled analyses. Fifth, the exclusion of non-English studies may have introduced language bias and underrepresentation of certain geographic regions. Sixth, female patients were underrepresented across included studies, limiting sex-specific insights. Lastly, long-term outcomes remain sparsely reported, and the lack of randomized controlled trials limits causal inference. Future prospective, standardized trials—ideally with diverse patient populations and harmonized data collection—are needed to further define the role of STAR in VT management. Standardization of STAR protocols and reporting practices is critical to advancing this field. Collaborative efforts are needed to develop robust frameworks for patient selection, target delineation, and outcome reporting. Future research should focus on:

1. Expanding STAR to earlier-stage VT patients, including those without prior catheter ablation failure.
2. Exploring particle therapy, particularly protons, for cases requiring enhanced OAR sparing.
3. Investigating advanced imaging and motion management techniques to improve precision.

4. Addressing geographic disparities by fostering international collaborations and improving access to radiation therapy technologies in underserved regions.

The findings of this review and pooled analysis underscore the transformative potential of STAR in VT management, while highlighting opportunities to refine and expand its application. Addressing the outlined challenges will be critical to maximizing STAR's clinical impact and ensuring equitable access to this life-saving technology.

In conclusion, STAR is poised to reshape VT management, bridging gaps in treatment options for patients unfit for catheter ablation. With continued collaboration, technological advancements, and equitable implementation, STAR holds the promise of evolving from an innovative alternative to an essential pillar of arrhythmia care.

**Acknowledgement**

This research is supported in part by the National Institutes of Health under Award Number R01CA272991, R01EB032680, R01DE033512, R37CA272755 and U54CA274513.

**Declaration of Generative AI and AI-assisted technologies in the writing process**

During the preparation of this work the authors used ChatGPT in order to enhance the readability of the manuscript. After using this tool, the authors reviewed and edited the content as needed and takes full responsibility for the content of the publication.

**Supplementary Materials**

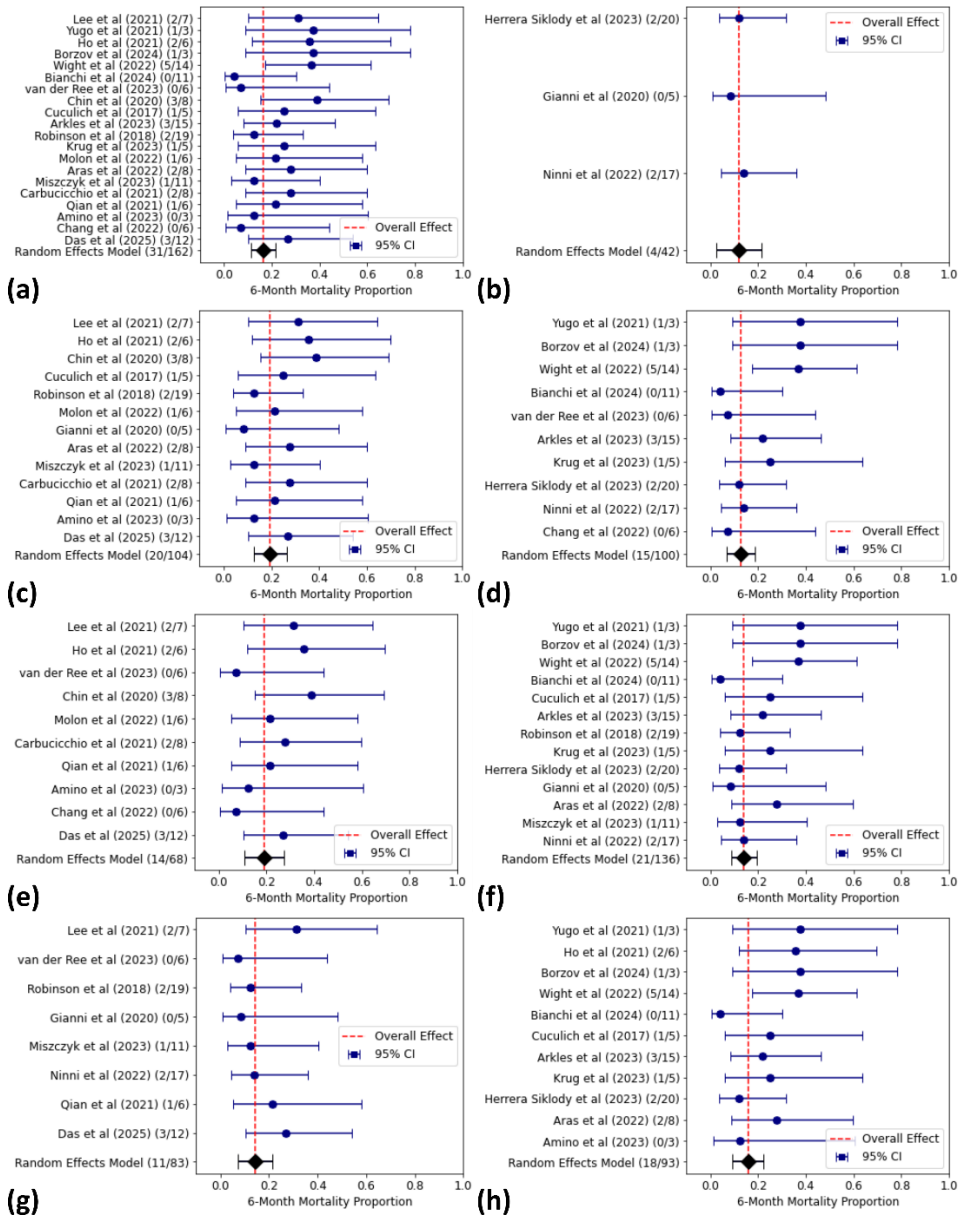

**Figure S1.** Forest plots depicting pooled mortality proportion 6 months following stereotactic arrhythmia radioablation (STAR), stratified by key subgroups. Each plot shows individual study estimates with 95% confidence intervals and an overall pooled estimate using a random-effects model. Studies are labeled with their reported VT reduction (%) and ordered chronologically. Subgroup comparisons include: **(A–B)** treatment modality — LINAC-based STAR **(A)** vs. CyberKnife **(B)**, **(C–D)** baseline left ventricular ejection fraction (LVEF) ≤ median **(C)** vs. > median **(D)**, **(E–F)** patient age ≤ median **(E)** vs. > median **(F)**, and **(G–H)** underlying cardiomyopathy — ischemic (ICM) **(G)** vs. non-ischemic (NICM) **(H)**.

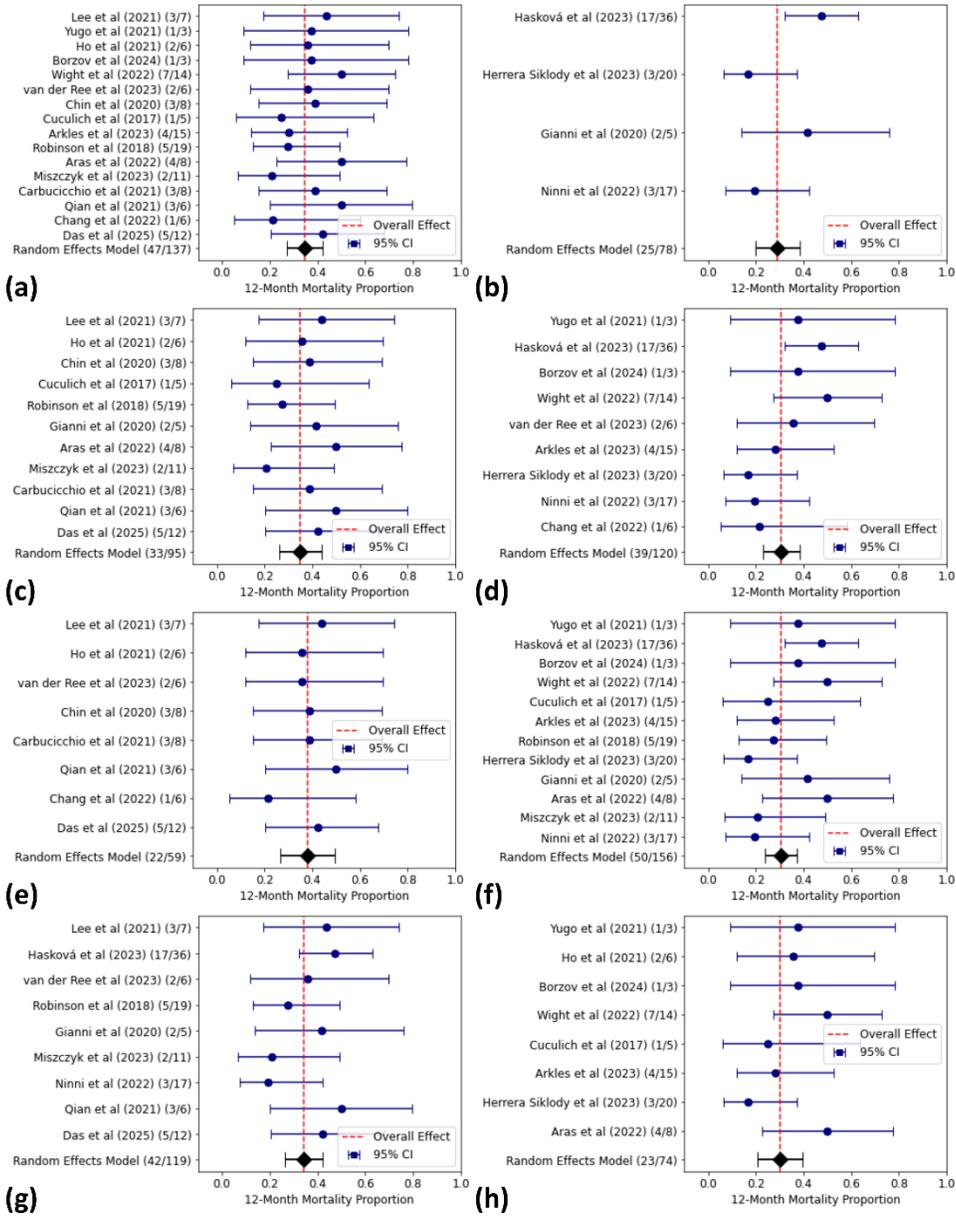

**Figure S2.** Forest plots depicting pooled mortality proportion 12 months following stereotactic arrhythmia radioablation (STAR), stratified by key subgroups. Each plot shows individual study estimates with 95% confidence intervals and an overall pooled estimate using a random-effects model. Studies are labeled with their reported VT reduction (%) and ordered chronologically. Subgroup comparisons include: **(A–B)** treatment modality — LINAC-based STAR **(A)** vs. CyberKnife **(B)**, **(C–D)** baseline left ventricular ejection fraction (LVEF) ≤ median **(C)** vs. > median **(D)**, **(E–F)** patient age ≤ median **(E)** vs. > median **(F)**, and **(G–H)** underlying cardiomyopathy — ischemic (ICM) **(G)** vs. non-ischemic (NICM) **(H)**.

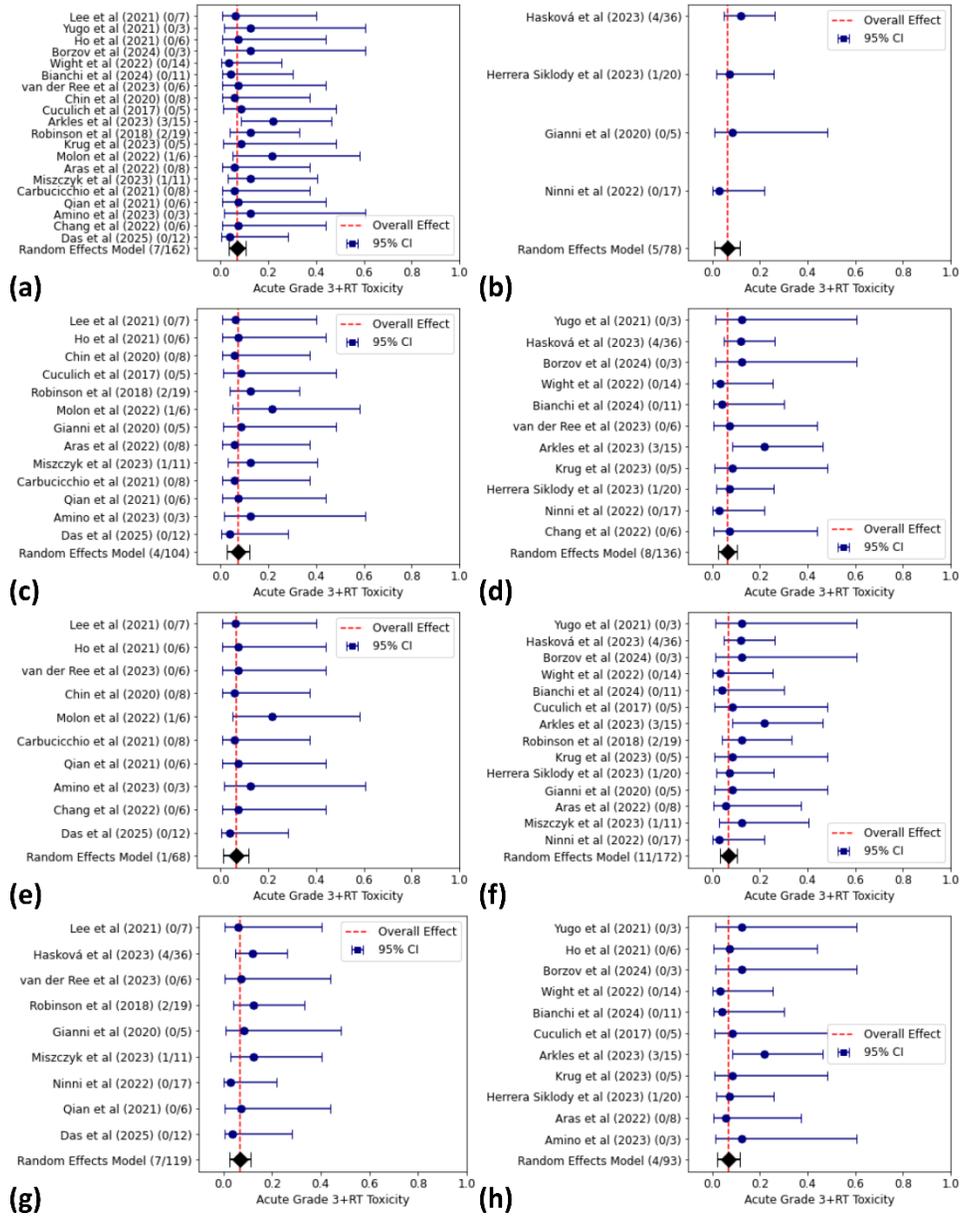

**Figure S3.** Forest plots depicting pooled acute adverse events following stereotactic arrhythmia radioablation (STAR), stratified by key subgroups. Each plot shows individual study estimates with 95% confidence intervals and an overall pooled estimate using a random-effects model. Studies are labeled with their reported VT reduction (%) and ordered chronologically. Subgroup comparisons include: **(A–B)** treatment modality — LINAC-based STAR **(A)** vs. CyberKnife **(B)**, **(C–D)** baseline left ventricular ejection fraction (LVEF) ≤ median **(C)** vs. > median **(D)**, **(E–F)** patient age ≤ median **(E)** vs. > median **(F)**, and **(G–H)** underlying cardiomyopathy — ischemic (ICM) **(G)** vs. non-ischemic (NICM) **(H)**.